\documentstyle[12pt]{article}

\catcode`\@=11 \@addtoreset{equation}{section}\catcode`\@=12
\newcommand{\be}{\begin{equation}}
\newcommand{\ee}{\end{equation}}
\newcommand{\ba}{\begin{eqnarray}}
\newcommand{\ea}{\end{eqnarray}}

\begin{document}
\thispagestyle{empty}

\begin{center}
               RUSSIAN GRAVITATIONAL SOCIETY\\
               INSTITUTE OF METROLOGICAL SERVICE \\
               CENTER OF GRAVITATION AND FUNDAMENTAL METROLOGY\\

\end{center}
\vskip 4ex
\begin{flushright}
                                         RGS-CSVR-001/96
                                         \\ gr-qc/xxxxx

 \end{flushright}
\vskip 15mm

\begin{center}
{\bf
 MULTIDIMENSIONAL INTEGRABLE VACUUM COSMOLOGY \\
              WITH TWO CURVATURES}

\vskip 5mm
{\bf  V. R. Gavrilov,
V. D. Ivashchuk and V. N. Melnikov }\\
\vskip 5mm

      {\em Center for Gravitation and Fundamental Metrology, VNIIMS,\\
     3-1 M.Ulyanovoy St., Moscow, 117313, Russia}\\
     e-mail: mel@cvsi.rc.ac.ru \\

\end{center}
\vskip 10mm

ABSTRACT

The vacuum cosmological model on the manifold
$R \times M_1 \times \ldots \times M_n$
describing the evolution of $n$ Einstein spaces of non-zero
curvatures is considered. For $n = 2$ the Einstein
equations are reduced to the Abel (ordinary differential)
equation and solved, when
$(N_1 = $dim $ M_1, N_2 = $ dim$ M_2) = (6,3), (5,5), (8,2)$.
The Kasner-like behaviour of the solutions near the
singularity $t_s \to +0$ is considered ($t_s$ is
synchronous time). The exceptional ("Milne-type") solutions
are obtained for arbitrary $n$. For $n=2$ these solutions
are attractors for other ones, when $t_s \to + \infty$. For dim $
M = 10, 11$ and  $3 \leq  n \leq 5$  certain two-parametric families of
solutions are obtained from $n=2$ ones using
"curvature-splitting" trick. In the case $n=2$, $(N_1, N_2)= (6,3)$
a family of non-singular solutions with the topology
$R^7 \times M_2$ is found.

\vskip 10mm


\vskip 30mm

\centerline{Moscow 1996}
\pagebreak

\setcounter{page}{1}

\pagebreak

%%%%%%%%%%%%%%%%%%%%%%%%%%%%%%%%%%%%%%%%%%%%%%%%%%%%%%%%%%%%%%%%%%
\section{Introduction}
\setcounter{equation}{0}
%%%%%%%%%%%%%%%%%%%%%%%%%%%%%%%%%%%%%%%%%%%%%%%%%%%%%%%%%%%%%%%%%%

Usually within multidimensional cosmology (see, for instance,
\cite{CD}-\cite{WP} and refs. therein) the space-time is considered
as a manifold
\begin{equation} %1.1
M = R \times M_{1} \times \ldots \times M_{n},
\end{equation}
where $R$ is the time axis, one of  the manifolds $M_1, \ldots, M_n$ is
interpreted as $3$-dimensional external space and the others as so-called
internal spaces. In most models $M_1, \ldots, M_n$ are considered as
spaces of constant Riemann curvature, or Einstein spaces. From the very
beginning it is supposed that the usual
three spatial dimensions and extra
spatial dimensions are on the same footing and this assumption is followed
from believing (of some physicists)  that the early Universe is
multidimensional. The separation of extra dimensions from the usual one is
attained in multidimensional cosmology by the so-called dynamical
compactification of internal spaces to unobservable sizes when their scale
factors decrease to the lengths of the Planck order ($10^{-33} cm$)
during the Universe evolution.

The study of  dynamical properties of the multidimensional Universe such
as, for instance, compactification of internal spaces or expanding of
the external space in detail demands exact solutions. However, in all
known integrable multidimensional cosmological models the
chain of the constant curvature spaces $M_1,\ldots,M_n$ may contain at most
one space of non-zero curvature  (see, for example, \cite{I}).
As far as we know, there are no  integrability conditions or explicit
integration methods for the models describing evolution of two or more
Einstein spaces with non-zero Ricci tensors. The aim of this paper is to
integrate the vacuum model on the manifold $R\times M_1\times M_2$, where
the Einstein spaces $M_1$ and $M_2$ have non-zero Ricci tensors.

This paper is organized as follows. In Sect. $2$ we describe the model
and get the equations of motion. We show, that their integrability by
quadratures is reduced to the integrability of some ordinary differential
equation. The latter appeared to be Abel's equation. In Sect.
$3$, we show, that for the models with  (dim$M_1, $dim$M_2) = (6,3),
(5,5), (8,2)$ the Abel equation belongs to the integrable class
described by Zaitsev and Polyanin \cite{ZP,PZ}  within discrete-group
analysis methods.  We integrate the (vacuum) Einstein equations for these
three models (see formula (3.38)) for non-exceptional case  and present
the trajectories of motion on the scale factors configuration plane.  In
Sect. $4$ the exceptional "Milne-type" solutions  (see (4.1)
and generalization for arbitrary $n$  in (4.2)) are presented.  The
Kasner-like behaviour near the singularity (for $t_s \to +0$) is
investigated in Sect. $5$.  In  Sect. $6$ certain two-parametric families
of solutions with $n$ curvatures for $3 \leq n \leq 5$ are presented. Some
non-singular solutions with topology $R^7 \times M_2$ are considered in
Sect. $7$.

\section{The model}
\par

At first, following our previous papers \cite{IM1}-\cite{Mel2}
we consider here the  general formalism for description of
multidimensional vacuum cosmological models. It is supposed, that
$D$-dimensional space-time manifold $M$ is defined by relation  (1.1).
The manifold $M$ is equipped with the
metric
\begin{equation} %2.1
g=-\exp[2{\gamma}(t)]dt \otimes dt +
\sum_{i=1}^{n}\exp[2{x^{i}}(t)]g^{(i)},
\end{equation}
where $\gamma(t)$ is an arbitrary function determining
the time $t$. It is supposed that the manifold
$M_i$ (see (1.1)) for $i=1,\ldots,n$ is the Einstein space of dimension
$N_i$ with the metric $g^{(i)}$, i.e.  \begin{equation} %2.2
{R_{m_{i}n_{i}}}[g^{(i)}] = \lambda_{i} g^{(i)}_{m_{i}n_{i}}, \ \
m_{i},n_{i}=1,\ldots ,N_{i},
\end{equation}
where $\lambda_i$ is constant.
(For  the manifold $M_i$ of
constant Riemann curvature $K_i$ the constant $\lambda_i$ reads:
$\lambda_i= K_i (N_i-1)$).

The non-zero components of the Ricci tensor for the metric
(2.1) are the following \cite{I}
\begin{equation} %2.3
R_0^0= \exp[-2\gamma]\sum_{i=1}^{n} N_{i}[ \ddot{x}^{i} - \dot{\gamma}
\dot{x}^{i} + (\dot{x}^{i})^{2}],
\end{equation}

\begin{equation} %2.4
R_{k_i}^{l_i}=\delta_{k_i}^{l_i}
\left( \lambda_{i} \exp[-2x^i] +
\left[\ddot{x}^{i}+ \dot{x}^{i} \left(\sum_{j=1}^{n}N_{j}\dot{x}^{j}-
\dot{\gamma}\right)\right] \exp[-2\gamma]\right),
\end{equation}
where indices $k_i$ and $l_i$ for $i=1,\ldots,n$ run over from
$1+\sum_{j=1}^{i-1}N_j$ to $\sum_{j=1}^{i}N_j$.

After the gauge fixing $\gamma = {F}(x^i)$
vacuum Einstein equations $R_B^A=0$ for $A,B=1,\ldots, D-1$ and
equation $R_0^0-\delta_0^0R/2=0$ may be presented as the Lagrange-Euler
equations and zero-energy constraint correspondingly obtained from
some Lagrangian $L(x^i,\dot{x}^i) = {L_{F}}(x^i,\dot{x}^i)$.
For the so-called harmonic time gauge
\begin{equation} %2.5
\gamma = \gamma_{0} \equiv \sum_{i=1}^{n} N_{i}x^{i}
\end{equation}
this Lagrangian has the following form \cite{IMZ}
\begin{equation} %2.6
L(x^i,\dot{x}^i)
=\frac{1}{2}\sum_{i,j=1}^{n}G_{ij}\dot{x}^{i}\dot{x}^{j}+
\frac{1}{2}\sum_{i=1}^{n}\lambda_{i}N_{i}\exp[-2x^{i}+2\gamma_{0}],
\end{equation}
where
\begin{equation}  %2.7
G_{ij}=N_{i}\delta_{ij}-N_{i}N_{j}
\end{equation}
are components of the minisuperspace metric.
The minisuperspace metric has
the pseudo-Euclidean signature $(-,+,\ldots,+)$ \cite{IM2,IMZ}.
(This is a well-known property of the
Hilbert-Einstein action \cite{W}.)

If all spaces $M_1,\ldots,M_n$ are Ricci-flat, i.e. $\lambda_i=0$
for $i=1,\ldots,n$, or only one of them has non-zero Ricci tensor,
then  equations of motion following from the Lagrangian (2.6) are
integrable and exact solutions are obtained in \cite{I}. But nearly
nothing is known about integrability of the models for two or more
Einstein spaces with non-zero Ricci tensor. Obtaining  exact solutions
for such models is connected with a general problem of integrability of
the so-called pseudo-Euclidean Toda-like systems \cite{I,IM3}. These
systems are described by the Lagrangian of the following form
\begin{equation} %2.8
L=\frac{1}{2} \sum_{a,b=0}^{n-1} \eta_{ab} \dot{z}^{a} \dot{z}^{b} -
\sum_{s=1}^{m} A^{(s)} \exp[ \sum_{a=0}^{n-1} u_a^s z^a],\ \
\end{equation}
where all $u^s_a$, $A^{(s)}$ are real constants and
$(\eta_{ab})=$diag$(-1,+1,\ldots,+1)$. The Lagrangian (2.6) may be reduced
to the Lagrangian (2.8) by some linear transformation of the configuration
variables $x^1,\ldots,x^n$.
It should be noted, that the
Lagrangian (2.8) describes both vacuum multidimensional cosmological models
or models with multicomponent perfect fluid source \cite{IM3,GIM}. The
well-known $4$-dimensional Bianchi-IX model (the so-called "mixmaster
model") is also described by the Lagrangian (2.8) with $n=3$, $m=6$ and
certain $u^s_a$, $A^{(s)}$ \cite{M}.

There are extensive literature devoted to the integrability of the
Euclidean Toda-like systems ($(\eta_{ab})=$diag($+1,\ldots,+1)$ in (2.8))
and methods of their explicit integration (see, for instance, \cite{T}),
but, as far as we know, the integrability of the pseudo-Euclidean systems
is not well studied yet. In our previous paper  \cite{GIM} we singled out
two integrable by quadratures classes of these systems and developed their
integration.  The first class contains the pseudo-Euclidean systems
trivially reducible to the Euclidean systems. It is easy to see that
under the conditions:  $u^s_0=0$ for $s=1,\ldots,m$ it follows from (2.8)
that $\dot{z}^0=$const. Then the Lagrangian (2.8) effectively
"loses" negative term in the kinetic energy, so for
$z^1, \ldots, z^{n-1}$ the Euclidean system
arises.  The second class appears provided vectors $u^1,\ldots, u^m \in
R^n$, where $u^s=(u^s_0,\ldots, u^s_{n-1})$, are collinear or form the
orthogonal set with respect to the symmetrical bilinear form on $R^n$ with
matrix $(\eta_{ab})$ (see for details  \cite{GIM}). But the models
describing evolution of two or more Einstein spaces with  non-zero Ricci
tensors are out of these classes and until now we had neither any
integrability conditions nor explicit integration methods for them. In
this paper we show that at least the models with $n=2$, i.e.
multidimensional vacuum cosmological models with metric (2.1) on the
manifold (1.1), where $M_1$ and $M_2$ are Einstein spaces of certain
dimensions with non-zero Ricci tensor, are integrable by quadratures, and
develop their integration procedure.

So, let us consider the Lagrangian (2.6) for $n=2$ and $\lambda_1$,
$\lambda_2 \neq 0$. After the following coordinate transformation
diagonalizing the minisuperspace metric:
\begin{eqnarray}  %2.9-10
&&x= x^1 -  x^2 + \frac{1}{2}
\ln{\left|\frac{\lambda_2 N_2}{\lambda_1 N_1}\right|}, \\
&&y= \alpha \beta \left(N_1 x^1 + N_2 x^2 \right),
\end{eqnarray}
where $\alpha$ and $\beta$ are defined by the relations
\begin{equation}  %2.11
\alpha = \sqrt{\frac{N_1 + N_2-1}{N_1 + N_2}}, \ \
\beta = \sqrt{\frac{N_1 + N_2}{N_1 N_2}},
\end{equation}
we obtain the following expression for the Lagrangian (2.6) in the new
variables $x$ and $y$
\begin{equation}  %2.12
L= \frac{1}{2 \beta^2} \left(\dot{x}^2-\dot{y}^2 \right)- V(x,y),
\end{equation}
where the potential $V(x,y)$ has the form
\begin{equation}  %2.13
V(x,y)= V_0 \exp(2 \alpha \beta^{-1} y)
[ {\rm sgn}(\lambda_1) \exp(2 \beta_1 x) +
{\rm sgn}(\lambda_2) \exp(2 \beta_2 x) ].
\end{equation}
Here we  denoted
\begin{equation} %2.14
V_0= - \frac{1}{2} \left|\lambda_1 N_1 \right|^{\beta_2}
\left|\lambda_2 N_2 \right|^{-\beta_1}, \ \
\beta_1  = - \frac{N_2}{N_1 + N_2}, \ \
\beta_2  =  \frac{N_1}{N_1 + N_2}.
\end{equation}
We note that $\beta_2 - \beta_1 = 1$.
The corresponding zero-energy constraint reads
\begin{equation} %2.15
E \equiv \frac{1}{2\beta^2} \left(\dot{x}^2- \dot{y}^2 \right)+
V(x,y)=0.
\end{equation}
The Lagrangian (2.12) leads to the following equations of motion
\begin{eqnarray} %2.16- 17
&&\ddot{x}= - 2 \beta^2 V_0
\exp(2 \alpha \beta^{-1} y)
[\beta_1 {\rm sgn}(\lambda_1) \exp(2 \beta_1 x) +
 \beta_2 {\rm sgn}(\lambda_2) \exp(2 \beta_2 x) ], \\
&&\ddot{y}= 2 \alpha \beta V(x,y).
\end{eqnarray}

It is easy to see, that the set of equations (2.16), (2.17) does not admit
solutions with $\dot{x}=\dot{y}= 0$ $\forall t$.
Indeed, such solutions are possible only if
$\lambda_1 \lambda_2 < 0$  and $\beta_1 = \beta_2$. But the latter
condition is impossible in the model under consideration (see (2.14)). So,
our model has no static solutions for any dimensions of $M_1$ and
$M_2$.

To integrate the equations of motion (2.16), (2.17) under the zero-energy
constraint (2.15) we consider at first the following procedure proposed in
\cite{BLP}. Using the relation
\begin{equation} %2.18
y''\equiv \frac{d^2y}{dx^2}=  \frac{\ddot{y}-\ddot{x}y'}{\dot{x}^2},
\end{equation}
equations of motion (2.16),(2.17) and the zero-energy constraint (2.15)
in the form
\begin{equation} %2.19
\dot{x}^2=
2 \beta^2 \frac{V(x,y)}{(\dot{y}/\dot{x})^2-1} =
2 \beta^2 \frac{V(x,y)}{(y')^2-1},
\end{equation}
we obtain the following ordinary differential equation of the second order
\begin{equation}  %2.20
y''= \left[(y')^2-1  \right]
\left\{ \frac{1}{2} (\beta_1 + \beta_2 + f(x)) y'
+ \frac{\alpha}{\beta} \right\},
\end{equation}
where
\begin{eqnarray} %2.21
{f}(x)= &&\tanh(x), \ \ \ {\rm if}\
\lambda_1 \lambda_2>0, \\
&&\coth(x),\ \ \ {\rm if}\  \lambda_1 \lambda_2<0.
\end{eqnarray}

After this procedure is done the following solutions may be lost:
$\dot{x} = \pm \dot{y}$  $\forall t$ and $\dot{x}=0$  $\forall t$.
It is easy to see, that under the zero-energy constraint the equations of
motion (2.16), (2.17)  have no solutions with $\dot{x}= \pm \dot{y}$
$\forall t$. The solutions with $\dot{x}=0$  $\forall t$ appear only
in the case $\lambda_1<0$ and $\lambda_2 <0$. So, by integration of the
equation (2.20) we obtain all possible trajectories on the configuration
plane $(x,y)$ except the trajectory
$x=\frac{1}{2}\ln[- \beta_1 /\beta_2]$ for
the model with $\lambda_1<0$ and $\lambda_2<0$. If the trajectories of
motion are known we may get the law of motion for each trajectory by
integration of the equation (2.19). Thus, the problem of the integrability
by quadratures of the equations (2.16), (2.17) under the constraint (2.15)
is reduced to the problem of the integrability of the equation (2.20). In
the paper \cite{BLP} the equation (2.20) was studied by qualitative
methods, but here we consider its explicit integration.

For the considered model we have from (2.11) and (2.14)
\begin{equation} %2.23
\beta_1 + \beta_2 =  \frac{N_1 - N_2}{N_1 + N_2}, \qquad
\frac{\alpha}{\beta} = \frac{R}{N_1 + N_2},
\end{equation}
where
\begin{equation}  %2.24
R = {R}(N_1, N_2) \equiv \sqrt{N_1 N_2 (N_1 + N_2 - 1)}.
\end{equation}
We also rewrite (2.21), (2.22) as follows
\begin{equation}  %2.25
{f}(x) = \frac{\exp(2x) - \varepsilon}{\exp(2x) + \varepsilon},
\end{equation}
where here and below
\begin{equation}   %2.26
\varepsilon = \varepsilon_1 \varepsilon_2, \ \
\varepsilon_i = {\rm sgn}(\lambda_i),
\end{equation}
$i =1,2$.

It should be noted, that the right side of the equation (2.20) does not
contain the unknown function $y$ due to the factorization of the potential
$V(x,y)$. So, for the function $z \equiv y'$ we have the first order
equation of the form
\begin{equation} %2.27
z'=f_0(x)+f_1(x)z+f_2(x)z^2+f_3(x)z^3.
\end{equation}
An equation of the form (2.27) is well-known as Abel's  equation (see,
for instance \cite{ZP,PZ}). There are no methods to integrate Abel's
equation with arbitrary functions $f_0(x)$, $f_1(x)$, $f_2(x)$ and
$f_3(x)$.

%%%%%%%%%%%%%%%%%%%%%%%%%%%%%%%%%%%%%%%%%%%%%%%%%%%%%%%%%%%%%%%%%%%%%%%%%%
%%%%%%%%%%%%%%%%%%%%%%%%%%%%%%%%%%%%%%%%%%%%%%%%%%%%%%%%%%%%%%%%%%%%%%%%%%%%
%%%%%%%%%%%%%%%%%%%%%%%%%%%%%%%%%%%%%%%%%%%%%%%%%%%%%%%%%%%%%%%%%%%%%%%%%%%

\section{Exact solutions}
\par

If any solution $z_0(x)$ to the Abel equation (2.27) is known, then by
the transformation \cite{PZ}
\begin{equation} %3.1
u= \frac{E(x)}{z-z_0},\ \
E(x)=\exp\left[
\int\left(3f_3(x)z_0^2+2f_2(x)z_0+f_1(x)\right)dx\right]
\end{equation}
it may be written as
\begin{equation}  %3.2
uu'+F_1(x)+F_2(x)u=0,
\end{equation}
where
\begin{equation}  %3.3
F_1(x)=f_3(x)(E(x))^2,\ \ \ F_2(x)=\left(3f_3(x)z_0+f_2(x)\right)E(x).
\end{equation}
For the new variable
\begin{equation}  %3.4
q=-\int F_2(x)dx
\end{equation}
the Abel equation has the so-called canonical form \cite{PZ}
\begin{equation}  %3.5
u\frac{du}{dq}-u=F(q),
\end{equation}
where we denoted
\begin{equation} %3.6
F(q)=\frac{F_1(x)}{F_2(x)}.
\end{equation}
The Abel equation in the form (3.5) was recently studied by
a discrete-group analysis methods
and a number of new integrable equations were
described \cite{ZP,PZ}.  In principle, the Abel equation (2.20) may be
written in the canonical form as its solutions $y'=\pm 1$ are known, but
some difficulty to obtain the function $F(q)$ in an explicit form for
arbitrary parameters $\alpha$, $\beta$ and $\beta_i$ arises. It is not
hard to verify, that if one of the two following conditions holds
\begin{equation}  %3.7
- \alpha = \beta (2 \beta_1 - \beta_2), \ \ \
 \alpha = \beta (2 \beta_2 - \beta_1),
\end{equation}
then the function $F(q)$ for the equation (2.20) may be obtained in
an explicit form by solving  some algebraic second order equation.
Using (2.11) and (2.14) we get corresponding relations for the dimensions
of $M_1$ and $M_2$.

\begin{equation}  %3.8
N_1=\frac{4N_2}{N_2-1},\ \ \ N_2=\frac{4N_1}{N_1-1}.
\end{equation}
It can be easily  seen that the first
condition holds in the following three cases:
i) $N_1=6$ and $N_2=3$; ii) $N_1=8$ and $N_2=2$,
iii) $N_1=5$ and $N_2=5$.
The second condition corresponds
to the inverse numbering of the spaces $M_1$ and $M_2$.

Under  one of the conditions (3.8) the Abel
equation (2.20) in the canonical form may be reduced to the Emden-Fowler
equation of the form:  $d^2Y/dX^2=$const$Y^{{\rm const}}$ \cite{PZ}, which
may be easily integrated.

\subsection{(1+6+3)-model}

First, let us  consider only the model with
$N_1=6$ and $N_2=3$. This $(1+6+3)$-model is of most interest, because the
$3$-dimensional space $M_2$ may be interpreted as our (external) space.
Omitting technical details we give at once the final Emden-Fowler equation
for this model. One may check that the Abel equation
(2.20) for $N_1=6$ and $N_2=3$ by the transformation \cite{PZ}
\begin{eqnarray} %3.9 -10 37
&&1+ \varepsilon \exp[2x]=
\frac{3X}{2Y}\frac{dY}{dX},\\
&&y'=1+\frac  {3\frac{X}{Y}\frac{dY}{dX}-2}
{1+\frac{X}{Y}\frac{dY}{dX}
\left[\frac{a}{Y}\left(\frac{dY}{dX}\right)^{-2}-1\right]},
\end{eqnarray}
where $a \neq 0$ is arbitrary constant, may be reduced to the following
Emden-Fowler equation
\begin{equation}  %3.11
\frac{d^2Y}{dX^2}= aY^{-2},
\end{equation}
which is well-known in classical mechanics as describing the
one-dimensional motion of a charged particle in the  field of the
(attractive or repulsive) Coulomb center.  We remind that
$\varepsilon = {\rm sgn}(\lambda_1 \lambda_2)$. So, by integrating
(3.11) and using the transformation (3.9), (3.10) we obtain the general
solution to the equation (2.20) for $N_1=6$ and $N_2=3$. In a parametrical
form the result looks as follows
\begin{eqnarray} %3.12-14
&& \varepsilon
\exp[2x]= {\Phi}(\tau, - \delta, C_1)
= \frac{3}{2} \delta \left(\tau^2 -
\delta \right) \left[ 1 + \tau \left( g(\tau, - \delta)+ C_1 \right)
\right] + \frac{1}{2}, \\ &&y'=\frac{\left(1+ 3\delta \tau^2 \right)
\exp[2x]+ \varepsilon} {\left(1- 3\delta\tau^2 \right) \exp[2x]+
\varepsilon}, \\
&&y=x-\frac{3}{2} \ln\left|\tau^2-\delta\right| +
C_2,
\end{eqnarray}
where
\begin{eqnarray}  %3.15-16
g(\tau,  - \delta)=
  &&\frac{1}{2} \ln\left|\frac{\tau-1}{\tau+1}\right|, \ \ \delta=+1, \\
= &&\arctan(\tau), \qquad \delta = -1.
\end{eqnarray}
By $\tau$ we denote the parameter (time), $C_1$ and $C_2$ are arbitrary
constants. Formulas (3.12), (3.13) together with $y'=\pm 1$ present the
general solution to the Abel equation (2.20) for $N_1=6$ and $N_2=3$.
This  may be also verified by a straightforward calculation
using the following relation
\begin{equation}  %3.17
\dot{\Phi} = - \frac{\delta}{\tau
(\tau^2 - \delta)} \left[ \left(1- 3 \delta \tau^2 \right) \Phi + 1.
\right]
\end{equation}
The function (3.12) is depicted in Figs. 1, 2 for different
values of $\delta$ and $C_1$. The limits  ${\Phi}(\tau) \to \infty$
and ${\Phi}(\tau) \to 0$ (for non-exeptional $C_1$) correspond
to Kasner-like behaviour near the singularity (see Sec. 5 below).

              {\bf Figs. 1, 2}

Using relations (3.12), (3.17),  zero-energy constraint (2.15) and
the expression for the potential (2.13) we obtain
the following relation between harmonic time and $\tau$-variable
\begin{equation} %3.18
\left(\frac{d\tau}{dt}\right)^2 = A (-\delta \varepsilon_2)
|\tau^2 - \delta|^{-2} \exp \left(4x + \frac{8}{3} C_2 \right),
\end{equation}
where
\begin{equation} %3.19
A =  \frac{1}{6} |6 \lambda_1|^{2/3}  |3 \lambda_2|^{1/3}.
\end{equation}
Thus, from (3.18) we obtain
\begin{equation} %3.20
\delta  = - \varepsilon_2 = - {\rm sgn}(\lambda_2).
\end{equation}

The transformation inverse to (2.9), (2.10)
for $N_1=6$, $N_2=3$ looks as follows
\begin{eqnarray} %3.21 -22 48 -49
&&x^1 =
\frac{1}{6} \left(2x + y -
\ln \left|\frac{\lambda_2}{2 \lambda_1}\right|\right),\\
&&x^2=
\frac{1}{6}\left(-4x+y+
2\ln \left|\frac{\lambda_2}{2\lambda_1} \right|\right).
\end{eqnarray}

Using  (2.5), (3.12), (3.14), (3.18)-(3.22) we get the following
relation for the metric (2.1) ($n = 2$)
\begin{equation} %3.23
g = c \left(f_1 f_2 \right)^{- \frac{1}{2}}
\{ - 2 f_1^{-2}  d\tau \otimes d\tau  +
f_2 |\lambda_1| g^{(1)} + |\lambda_2| g^{(2)} \}.
\end{equation}
Here  $c \neq 0$ is an arbitrary constant and
\begin{eqnarray} %3.24 - 25
f_1&=&{f_1}(\tau, \varepsilon_2) =
\left|\tau^2 + \varepsilon_2 \right|, \\
f_2&=&{f_2}(\tau, \varepsilon_1, \varepsilon_2, C_1) =
2 \varepsilon_1 \varepsilon_2  {\Phi}(\tau, \varepsilon_2, C_1)
\nonumber \\
&=&- 3 \varepsilon_1 \left(\tau^2 + \varepsilon_2 \right)
\left[ 1 + \tau \left( g(\tau, \varepsilon_2)+ C_1 \right) \right]
+ \varepsilon_1 \varepsilon_2 > 0.
\end{eqnarray}
We recall that $\varepsilon_i = {\rm sgn}(\lambda_i)$, $i =1,2$.
It should be noted that although originally
$c > 0$ since $c = 2^{-1/2} A^{-1} \exp(C_2/3)$, the negative $c $ also
gives us the solution to vacuum Einstein equations.
The trajectories on the plane of scale factors $a_i  > 0, i=1,2$,
corresponding to the solution (3.23), where
$$a_1^2 = c \left(f_1 f_2 \right)^{- \frac{1}{2}}f_2 |\lambda_1|,
\qquad
a_2^2 =
c \left(f_1 f_2 \right)^{- \frac{1}{2}} |\lambda_2|,$$
are depicted in Figs. 3-5
(we put $c= |\lambda_1| = |\lambda_2| =1$).  The points
$\tau_i = {\tau_i}(C_1)$, $i=1,2,3$, are zeros of $f_2$.

                {\bf   Figs. 3 - 5 }

\subsection{The (1+8+2)- and (1+5+5)-models}

Some modification of the Anzatz (3.12)-(3.14) may be used for
obtaining the solutions to the Abel equation for other two cases
$(N_1, N_2) = (8,2), (5,5)$. The solutions read
\begin{eqnarray} %3.26 -28
&& \varepsilon \exp[2x]= {\Phi_k}(\tau, - \delta, C_1) \\
&&y'=\frac{\left(1+ k \delta \tau^2 \right) \exp[2x]+ \varepsilon}
{\left(1- k \delta \tau^2 \right) \exp[2x]+
\varepsilon}, \\
&&y = x - \frac{k}{2} \ln\left|\tau^2-\delta \right| + C_2,
\end{eqnarray}
where
\begin{equation} %3.29
 k = (N_1 + N_2)/N_2 = 3, \ 2, \ 5  \qquad
 {\rm  for} \ (N_1, N_2) = (6,3), \ (5,5), \ (8,2),
\end{equation}
 respectively and functions $\Phi_k  = {\Phi_k}(\tau, - \delta, C_1)$
satisfy to the equations
\begin{equation}  %3.30
\dot{\Phi}_k = - \frac{\delta}{\tau
(\tau^2 - \delta)} \left[ \left(1- k \delta \tau^2 \right) \Phi_k + 1
\right].
\end{equation}
The solutions to (3.30) are the following: $\Phi_3 = \Phi$ (see (3.12)) and
\begin{eqnarray} %3.31 -32
&&{\Phi_2}(\tau, - \delta, C_1) = 1 + 2 \delta \left(\tau^2 - \delta
\right) + C_1 \delta \tau \left| \tau^2 - \delta \right|^{\frac{1}{2}}, \\
&&{\Phi_5}(\tau, - \delta, C_1)  = \frac{1}{4} -
\frac{5}{8} \left( \tau^2 - \delta \right)
\left[ 3 \tau^2 - 2 \delta  + 3 \tau \left( \tau^2 - \delta \right)
\left( g(\tau, - \delta)+ C_1 \right) \right].
\end{eqnarray}

Using relations (3.26), (3.30) zero-energy constraint (2.15) and
the expression for potential (2.13) we obtain
the following relation between harmonic time and $\tau$-variable
\begin{equation} %3.33
\left(\frac{d\tau}{dt}\right)^2 = A (-\delta \varepsilon_2)
|\tau^2 - \delta|^{1-k} \exp \left[4x +
2 \left(1 + \frac{1}{k} \right) C_2 \right]
\end{equation}
where here
\begin{equation} %3.34
A =  \frac{1}{N_1} |N_1 \lambda_1|^{1- \frac{1}{k}}
|N_2 \lambda_2|^{\frac{1}{k}}.
\end{equation}
>From (3.33) we obtain the relation (3.20).

The transformation inverse to (2.9), (2.10)
is the following
\begin{eqnarray} %3.35 -36
&&x^1 =
\frac{N_2}{N_1 + N_2} \left( x  -  \frac{1}{2}
\ln \left|\frac{\lambda_2 N_2}{ \lambda_1 N_1} \right| \right)
+ \frac{q y}{N_1 + N_2},       \\
&&x^2 =
- \frac{N_1}{N_1 + N_2} \left( x  -  \frac{1}{2}
\ln \left|\frac{\lambda_2 N_2}{ \lambda_1 N_1} \right| \right)
+ \frac{q y}{N_1 + N_2},
\end{eqnarray}
where
\begin{equation} %3.37
q = (\alpha \beta)^{-1} = \sqrt{\frac{N_1 N_2}{N_1 + N_2 -1}}.
\end{equation}

Using  (2.5), (3.20), (3.26), (3.28), (3.33)-(3.37) we get the
relation for the metric (2.1) ($n = 2$)
\begin{equation} %3.38
g = c \left(f_1 f_2 \right)^{- r}
\{ - \frac{N_1}{N_2} f_1^{-2}  d\tau \otimes d\tau  +
f_2 |\lambda_1| g^{(1)} + |\lambda_2| g^{(2)} \}.
\end{equation}
Here  $c \neq 0$,
\begin{equation} %3.39
 r = q / N_2 = 2 - q = \frac{1}{2}, \ \frac{1}{3}, \ \frac{2}{3}
\end{equation}
for $(N_1,N_2) = (6,3), (5,5), (8,2)$ respectively
, $f_1 = {f_1}(\tau, \varepsilon_2)$ is defined in (3.24) and
\begin{equation} %3.40  67
f_2 = {f_2}(\tau, \varepsilon_1, \varepsilon_2, C_1,N_1,N_2) =
\frac{N_1}{N_2} \varepsilon_1 \varepsilon_2
{\Phi_k}(\tau, \varepsilon_2, C_1),
\end{equation}
where functions $\Phi_k$ are defined in (3.12), (3.31) and (3.32).

\section{Exceptional solution for negative curvatures}

Now, we consider the exceptional solution with $\dot{x}=0$. This
solution takes place, when $\lambda_1$, $\lambda_2<0$, and has the
following form
\begin{equation} %4.1
g=-dt_s\otimes dt_s + \sum^{2}_{i=1}\frac{\lambda_i}{(2-D)}t^2_s g^{(i)},
\end{equation}
where $D=1+ N_1+ N_2$. Here the synchronous-time parametrization is used.
Formula (4.1) may be obtained by solving eqs. (2.15)-(2.17) and
performing the coordinate transformation (2.9), (2.10) and
appropriate time reparametrization.

Moreover, it is not difficult to verify that the
metric
\begin{equation} %4.2
g=-dt_s\otimes dt_s + \sum^{n}_{i=1} \frac{\lambda_i}{(2-D)}t^2_s g^{(i)},
\end{equation}
defined on the manifold (1.1),
where $(M_i,g^{(i)})$ are Einstein spaces satisfying (2.2)
($\lambda_i \neq 0$) and dim$M=D$, is a solution to the vacuum Einstein
equations (or, equivalently, is Ricci-flat). (See (2.3), (2.4)).

Remark 1. We note that the metric
\begin{equation} %4.3
g^*=\sum_{i=1}^{n}\frac{\lambda_i}{(2-D)}g^{(i)}
\end{equation}
defined on the manifold $M_{1} \times \ldots \times M_{n}$
satisfies the same relation $R_{mn}[g^*]=-(D-2)g^*_{mn}$
as the metric $g(H^{D-1})$ of $(D-1)$-dimensional Lobachevsky
space. We also note that the metric
\begin{equation} %4.4
\eta = -dt_s \otimes dt_s + t_s^2 g(H^{D-1})
\end{equation}
is flat. For $D=3$ (4.4) coincides with the well-known Milne solution
\cite{Mi}. So,
the metric (4.2) may be called a "quasi-flat" or "Milne-like"  metric.

Remark 2. It may be proved that, when $(M_i,g^{(i)})$
are spaces of constant (non-zero) curvature, $i=1,..., n$, and
$n \geq 2$, the metric (4.2)  has a divergent Riemann-tensor
squared as $t_s \to +0$
\begin{equation} %4.5
I[g]=R_{MNPQ}[g]R^{MNPQ}[g] \to +\infty.
\end{equation}
In this case the solution (4.2) is singular. The relation (4.5)
follows from the formula
\begin{equation} %4.6
I[g]= {\cal A} t_s^{-4},
\end{equation}
where
\begin{equation}  %4.7
{\cal A} =(D-2)^2\sum_{i=1}^{n}\frac{I[g^{(i)}]}{\lambda_i^2}-2(D-1)(D-2).
\end{equation}
(see \cite{IM4,IM6}). For spaces of constant negative curvature
\begin{equation}  %4.8
{\cal A}=2(D-2)^2\left( \sum_{i=1}^{n}
\frac{N_i}{N_i-1}- \frac{D-1}{D-2} \right) > 0,
\end{equation}
if $n \geq 2$.

{\bf Isotropization}. Now, we show that the solution  (4.1) is an
attractor for  solutions (3.38) with  $\varepsilon_1 < 0$,
$\varepsilon_2 < 0$ and $c > 0$, as $\tau \to 1$ (or $t_s \to +\infty$).
Using the relation $f_2(\tau)\to 1$ as $\tau \to 1$, we get the
asymptotical formula for the synchronous-time variable $t_s$ ($c > 0$)
\begin{equation}
c^{1/2}t_{s} \sim
\frac{\sqrt{N_1/N_2}}{r2^{r/2}}|\tau-1|^{-r/2} \ {\rm as}\  \tau \to 1.
\end{equation}
Hence
\begin{equation}  %4.10
(f_1 f_2)^{-r}\sim\frac{r^2N_2}{N_1}t_s^2= \frac{t_s^2}{D-2}\
{\rm as}\  \tau \to 1.
\end{equation}
Thus, we obtain the relation for the metric
\begin{equation}  %4.11
g=-dt_s\otimes dt_s +
b_1(t_s)|\lambda_1|g^{(1)}+b_2(t_s)|\lambda_2|g^{(2)},
\end{equation}
where
\begin{equation} %4.12
b_i(t_s) \sim \frac{t_s^2}{D-2} \ \ {\rm as} \
t_s \to + \infty.
\end{equation}
%%%%%%%%%%%%%%%%%%%%%%%%%%%%%%%%%%%%%%%%%%%%%%%%%%%%%%%%%%%%%%%%%%%%%%%%%%%%

\section{Kasner-like behaviour for $t_s \to +0$}

Let us consider the solutions (3.38) written in the
synchronous-time parametrization
\begin{equation} %5.1
g=-dt_s\otimes dt_s + a_1^2(t_s)g^{(1)}+ a_2^2(t_s)g^{(2)},
\end{equation}
where
\begin{equation}  %5.2
t_s=\pm \int
^{\tau}_{\tau_0}d\tau^{\prime}(f_1f_2)^{-r/2} f_1^{-1}
\left(c \frac{N_1}{N_2}\right)^{1/2}+t_{0s},
\end{equation}
($f_i=f_i(\tau^{\prime}))$.
As follows from \cite{IM4} almost all solutions for the considered model
should have a Kasner-like behaviour for small $t_s$, i.e.
\begin{equation}  %5.3
a_i(t_s)\sim A_i t_s^{\alpha_i} \ \ {\rm as} \ t_s \to +0,
\end{equation}
where $A_i>0$ are constants and the Kasner parameters $\alpha_i$
satisfy the relations
\begin{equation} %5.4
\sum^{2}_{i=1} N_i\alpha_i=\sum^{2}_{i=1}N_i\alpha_i^2=1.
\end{equation}
Here a suitable choice of the constant $t_{0s}=t_{0s}(\tau_0)$ and
the sign in (5.2), such that $v= a_1^{N_1}a_2^{N_2}\to +0$
corresponds to  $t_s \to +0$, is assumed. We note, that although  in
\cite{IM4} models with minisuperspace dimensions  $n \geq 3$ were
considered, the results  are also applicable for the case $n=2$.

Solving the equations (5.4) we get \cite{BIMZ}
\begin{equation} %5.5
\alpha_1=\frac{N_1 \pm R}{(N_1+N_2)N_1}, \ \
\alpha_2=\frac{N_2 \mp R}{(N_1+N_2)N_2},
\end{equation}
where $R=R(N_1,N_2)$  is defined in (2.24).
For  $(N_1,N_2)=(6,3),(5,5), (8,2)$ $R$ is integer
\begin{equation} %5.6
R=N_1+2N_2=12, \ 15, \ 12
\end{equation}
correspondingly.
We obtain from (5.5)
\begin{eqnarray} %5.7
&&(\alpha_1, \alpha_2)_{+}= \left(\frac{1}{3},-\frac{1}{3}\right),
\ \left(\frac{2}{5},-\frac{1}{5}\right),
\ \left(\frac{1}{4},-\frac{1}{2}\right), \\
&&(\alpha_1,\alpha_2)_{-}= \left(-\frac{1}{9},\frac{5}{9} \right),
\ \left(-\frac{1}{5},\frac{2}{5} \right),
\ \left(-\frac{1}{20},\frac{7}{10}\right),
\end{eqnarray}
for $(N_1,N_2)=(6,3),(5,5), (8,2)$ respectively.
Here $(\alpha_1,\alpha_2)_{\pm}$ corresponds to "$\pm$" in (5.5).

Now, we return to the metric (3.38) written in the synchronous-time
form (5.1). The Kasner-like behaviour (5.3) with  $\alpha_i$ from (5.7)
takes place if $\tau \to \tau_{*}$, where $f_2(\tau_{*})=0$  and
$f_2^{\prime}(\tau_{*})\neq 0$. Indeed, from (5.2) we get
\begin{equation}  %5.9
t_s \sim {\rm const}|\tau-\tau_{*}|^{1- \frac{r}{2}} \ \
{\rm as} \ \tau \to \tau_{*}
\end{equation}
($r$ is defined in (3.39)) and using (3.38) we
obtain the Kasner-like behaviour (5.3) with the parameters
\begin{equation}   %5.10
\alpha_1= \frac{1-r}{2-r},\ \  \alpha_2=\frac{-r}{2-r},
\end{equation}
coinciding with those from eq. (5.7)

The Kasner-like asymptotes (5.3) with  parameters
from (5.8) take place, when $\tau \to \pm \infty$ , for
non-exceptional values of the constant $C_1$:
\begin{eqnarray}  %5.11 -12
&&C_1\neq 0,\ {\rm if}\ \varepsilon_2= -1 \
{\rm and} \ C_1 \neq \mp \frac{\pi}{2},\ {\rm if}\
\varepsilon_2=+1,\ {\rm for} \ (N_1,N_2)=(6,3),(8,2);\\
&&C_1\neq\pm 2 \ {\rm for} \  (N_1,N_2)=(5,5).
\end{eqnarray}
In this case
\begin{equation} %5.13
f_2(\tau)\sim B_{\pm}|\tau|^{k} \ \ {\rm as} \ \tau \to \pm \infty,
\end{equation}
where $B_{\pm}$ are constants (integer $k$ is defined in (3.29))
and
\begin{equation}  %5.14
t_s \sim {\rm const}|\tau|^{-1- \frac{s}{2}} \ \
{\rm as}\  \tau \to \pm \infty.
\end{equation}
Here $s = (k+2)r$. From (3.38), (5.13) and (5.14) we get the
Kasner-like behaviour (5.3) with  parameters
\begin{equation}  %5.15
\alpha_1=\frac{s-k}{2+s}, \ \
\alpha_2=\frac{s}{2+s},
\end{equation}
coinciding with (5.8).

According to the results of \cite{IM6}
the Riemann tensor squared for the solutions with
asymptotically Kasner-like behaviour  for $t_s \to +0$
is divergent
\begin{equation}  %5.16
I[g] = R_{MNPQ}[g] R^{MNPQ}[g]  \to + \infty
\end{equation}
as $t_s \to +0$ for fixed $(x_1, x_2) \in M_1 \times M_2$.
If $I[g^{(i)}] \geq c_i$ for some $c_i$, i = 1, 2, then the relation
(5.16) takes place uniformly on  $(x_1, x_2) \in M_1 \times M_2$
(see Theorem in \cite{IM6}).

%%%%%%%%%%%%%%%%%%%%%%%%%%%%%%%%%%%%%%%%%%%%%%%%%%%%%%%%%%%%%%%%%%%%%%%%%%%%
%%%%%%%%%%%%%%%%%%%%%%%%%%%%%%%%%%%%%%%%%%%%%%%%%%%%%%%%%%%%%%%%%%%%%%%%%%%%%
%%%%%%%%%%%%%%%%%%%%%%%%%%%%%%%%%%%%%%%%%%%%%%%%%%%%%%%%%%%%%%%%%%%%%%%%%%%%

\section{Special solutions with $n \leq 5$ curvatures}

The obtained above solutions may be used for generating
some special (two-parametric) classes  of the solutions to
vacuum Einstein equations. This may be done using the
"curvature-splitting" method described below.

Let us consider a set of $k$  Einstein manifolds $({\cal M}_i, h^{(i)})$
with non-zero curvature, i.e.
\begin{equation} %6.1
{\rm Ric}(h^{(i)})= \mu_i h^{(i)},
\end{equation}
where $\mu_i \neq 0$  is a real constant, $i = 1, ..., k$. Here and below
we denote by Ric$(h)$ the Ricci-tensor corresponding to the
metric $h$. Let $\mu \neq 0$ be a real number. Then
\begin{equation} %6.2
h= \sum_{i=1}^{k} \frac{\mu_i}{\mu} h^{(i)}
\end{equation}
is an Einstein metric, (correctly) defined on the manifold
\begin{equation} %6.3
{\cal M} =  {\cal M}_1 \times \ldots \times {\cal M}_k
\end{equation}
and satisfying
\begin{equation} %6.4
{\rm Ric}(h)= \mu h.
\end{equation}
Indeed,
\begin{equation} %6.5
{\rm Ric}(h)= \sum_{i=1}^{k} {\rm Ric}(\frac{\mu_i}{\mu}h^{(i)})=
\sum_{i=1}^{k} {\rm  Ric}(h^{(i)}) = \sum_{i=1}^{k}
\mu_i h^{(i)}= \mu h.
\end{equation}

Remark 3. In (6.2) (like in (2.1)) we identify the metric $h^{(i)}$  on
${\cal M}_i$ with its canonical extension to the manifold ${\cal M}$
(6.3). It is more correct  to write (instead of (6.2))
\begin{equation}   %6.6
h= \sum_{i=1}^{k}\frac{\mu_i}{\mu} p_i^* h^{(i)},
\end{equation}
where $p_i : {\cal M} \rightarrow {\cal M}_i$ is the canonical projection.
Analogously, Ric$(h^{(i)})$ in
(6.5)  should be understood as $p^*_i {\rm Ric}(h^{(i)})$ etc.

Now, using the suggested trick, we may consider the following Einstein
spaces $(M_i, g^{(i)})$ $(i = 1, 2)$ in (3.38):
\begin{eqnarray}  %6.7 - 6.8
&&g^{(i)}= \sum^{n_i}_{j=1}\frac{\lambda_{ij}}{\lambda_{i}}g^{(ij)},\\
&&M_i=M_{i1} \times \ldots \times M_{in_i},
\end{eqnarray}
where $(M_{ij},g^{(ij)})$ are Einstein spaces of non-zero curvature
\begin{equation}  %6.9
{\rm Ric}(g^{(ij)})=\lambda_{ij}g^{(ij)},\ \ \lambda_{ij} \neq 0,
\end{equation}
$j=1, \ldots, n_{i}$; $i=1,2$.
Clearly that
\begin{equation}  %6.10
N_i= \sum_{j=1}^{n_i} N_{ij}
\end{equation}
where
\begin{equation}  %6.11
N_{ij}= {\rm dim}M_{ij} >1,
\end{equation}
$j=1, \ldots n_{i}$; $i=1,2$. It follows from (6.10), (6.11) that
\begin{equation}  %6.12
1 \leq n_{i} \leq \left[ \frac{N_i}{2} \right],
\end{equation}
$i=1,2$; where  $[x]$ denotes the integer part of $x$.

Substituting (6.7)  into (3.38)  we get the following solutions
to (vacuum) Einstein equations:
\begin{equation}    %6.13
g= c (f_1f_2)^{-r}
\left\{-\frac{N_1}{N_2}f_1^{-2}d\tau\otimes d\tau+
\sum_{j=1}^{n_1} f_2 \varepsilon_1 \lambda_{1j}g^{(1j)}+
\sum_{l=1}^{n_2} \varepsilon_2 \lambda_{2l}g^{(2l)} \right\},
\end{equation}
defined on the manifold
\begin{equation}   %6.14
M= R \times M_{11} \times \ldots \times M_{1n_1}\times M_{21}\times
\ldots \times M_{2n_2}.
\end{equation}
The solution (6.13), (6.14) describes the evolution of $n=n_1+n_2$
Einstein spaces $(M_{ij}, g^{(ij)})$, satisfying  (6.9)-(6.12).
Here $(N_1,N_2)=(6,3),(5,5),(8,2)$; $c \neq 0$, $\varepsilon_i=\pm 1$;
$r=r(N_1,N_2)$ and $f_i =
{f_i}(\tau)$
are defined in subsections 3.1 and 3.2, $i = 1, 2$.

The relations (6.13), (6.14)  give us
\begin{equation}   %6.15
(1+4+2+3), \ (1+3+3+3), \ (1+2+2+2+3)
\end{equation}
solutions for $(N_1,N_2)=(6,3)$;
\begin{equation}   %6.16
(1+5+3+2), \ (1+3+2+3+2)
\end{equation}
solutions for $(N_1,N_2)=(5,5)$ and
\begin{eqnarray}   %6.17
(1+6+2+2), \ (1+5+3+2), \ (1+4+4+2),  \\ \nonumber
(1+4+2+2+2), \ (1+3+3+2+2), \ (1+2+2+2+2+2)
\end{eqnarray}
solutions for  $(N_1,N_2)=(8,2)$.

In the last case ($(8,2)$) we have two families (of solutions) of the same
type as in (6.16). (The corresponding solutions from (6.16) and (6.17)
seem to be different.) Thus, here we obtained some special
two-parametrical families of solutions to the problem of cosmological
evolution of $n$ curvatures for $n= 3, 4, 5$.
%%%%%%%%%%%%%%%%%%%%%%%%%%%%%%%%%%%%%%%%%%%%%%%%%%%%%%%%%%%%%%%%%%%%%%%%%%%%%
%%%%%%%%%%%%%%%%%%%%%%%%%%%%%%%%%%%%%%%%%%%%%%%%%%%%%%%%%%%%%%%%%%%%%%%%%%%%%
%%%%%%%%%%%%%%%%%%%%%%%%%%%%%%%%%%%%%%%%%%%%%%%%%%%%%%%%%%%%%%%%%%%%%%%%%%%%%
\section{Non-singular solutions}

Here we show that there exist non-singular solutions
among the considered ones. Let us restrict ourselves to the
$(1+6+3)$-case, with two negative curvatures $\varepsilon_1=
\varepsilon_2=-1$ and $C_1=0$. In this case the functions $g(\tau)$,
\begin{equation} %7.1
f_2(\tau)=3(\tau^2-1) \left[ 1 +  \frac{1}{2} \tau
\ln \left|\frac{\tau-1}{\tau+1}\right|\right]+1
\end{equation}
and $f_1(\tau)f_2(\tau)$  are holomorphic in
${\bar \Omega}=\Omega \cup \{+\infty\}$,
where $\Omega = C \setminus[-1,1]$
(${\bar \Omega}$ is the complex Riemann sphere with a cut),
\begin{eqnarray} %7.2 -4
&&g(\tau)=- \sum_{k=0}^{\infty}\frac{1}{2k+1}\tau^{-2k-1},\\
&&f_2(\tau)=\frac{2}{5\tau^2}+\sum_{k=2}^{\infty}f_{2,k}\tau^{-2k},\\
&&f_1(\tau)f_2(\tau)=\frac{2}{5}+\sum_{k=1}^{\infty}f_{,k}\tau^{-2k},
\end{eqnarray}
$|\tau|>1$.

Now we introduce a new time variable defined by the relation
\begin{equation}    %7.5
\ln{\frac{\rho}{\rho_0}}=-\int^{\tau}_{\tau_0}d\tau^{\prime}
\sqrt{\frac{2}{5f_1^2f_2}},
\end{equation}
where $\tau_0>1$, $\rho_0>0$. We may rewrite (7.5) as
\begin{equation}   %7.6
\rho=\frac{\rho_0\tau_0}{\tau}\exp[-I(\tau_0,\tau)],
\end{equation}
where
\begin{equation}  %7.7
I(\tau_0,\tau)=\int^{\tau}_{\tau_0}d\tau^{\prime}
\left[\sqrt{\frac{2}{5f_1^2f_2}}- \frac{1}{\tau}\right].
\end{equation}
The function (7.7) may be analytically continued to the
neighbourhood of $\infty$, i.e. in $\{|\tau|>T\}\cup\{\infty\}$
for some $T$, and, clearly that
\begin{equation} %7.8
I(\tau_0,\tau)=I(\tau_0,+\infty)+
\sum_{k=1}^{\infty}I_k \left(\frac{1}{\tau}\right)^{2k}.
\end{equation}
We put
\begin{equation} %7.9
\rho_0 \tau_0 \exp\left[-I(\tau_0,+\infty)\right]=1.
\end{equation}
Then $\rho \sim\frac{1}{\tau}$ as $\tau \to \infty$ . The function
$\rho= \rho(\tau)$ (7.6) is smooth and monotonically decreasing on
$(1,+ \infty)$  and may be analytically continued to
$\{|\tau|>T\} \cup \{\infty\}$, where
\begin{equation} %7.10
\rho(\tau)=\frac{1}{\tau}
\left[1+\sum_{k=1}^{\infty}\rho_k
\left(\frac{1}{\tau}\right)^{2k}\right].
\end{equation}
For $\tau \to 1$ from (7.5) we get
\begin{equation} %7.11
\rho(\tau) \sim{\rm const}(\tau-1)^{-1/\sqrt{10}}.
\end{equation}
Now, we put also
\begin{equation} %7.12
M_1=S^6,\ \ g^{(1)}= -g(S^6)=-d\Omega^2_6,
\end{equation}
where $d \Omega^2_6$ is the standard metric on $6$-dimensional sphere,
normalized by the condition
\begin{equation}  %7.13
{\rm Ric}\left(d\Omega^2_6\right)= 5 d\Omega^2_6.
\end{equation}
For this special case we get from (3.23), (7.5), (7.12), (7.13)
\begin{equation}  %7.14
g= c F_1(\rho)\left\{
-F_2(\rho)\left[d\rho\otimes d\rho+\rho^2d\Omega^2_6\right]
+|\lambda_2|g^{(2)}\right\},
\end{equation}
where
\begin{eqnarray} %7.15 -16
&&F_1(\rho)=\left[f_1(\tau(\rho))f_2(\tau(\rho))\right]^{-1/2},\\
&&F_2(\rho)=\frac{5f_2(\tau(\rho))}{\rho^2}.
\end{eqnarray}
Here the inverse function $\tau=\tau(\rho)$  is smooth for
$\rho\in(0,+\infty)$ and may be analytically continued to the domain
$\{|\rho|<\delta\}$  for some $\delta>0$. The same is valid for
the functions (7.15), (7.16) (see (2.3) and (2.10)). We have also for
$|\rho|<\delta$: $F_i(\rho)=F_i(-\rho)$, $i=1,2$,
$\tau(-\rho)=-\tau(\rho)$
and
\begin{eqnarray} %7.17 -18
&&F_1(\rho)=\left(\frac{2}{5}\right)^{-1/2}
\left[1+\sum_{k=1}^{\infty}F_{1,k}\rho^{2k}\right],\\
&&F_2(\rho)=2+ \sum_{k=1}^{\infty}F_{2,k}\rho^{2k}
\end{eqnarray}
(see (7.3), (7.4), (7.10)).

There exist functions $\Phi_i(w)$, $i=1,2$, defined on
$\{|w|<\delta^2\} \cup (0,+\infty)$, satisfying
\begin{equation} %7.19
\Phi_i(\rho^2)=F_i(\rho),
\end{equation}
holomorphic in $\{|w| < \delta^2\}$ and smooth on $(0,+\infty)$.
In terms of these functions the solution (7.14) may written as
\begin{equation} %7.20
g=c \Phi_1(|{\vec x}|^2)\left\{
-\Phi_2(|{\vec x}|^2) \sum_{i=1}^{7}dx^i \otimes dx^i+
|\lambda_2|g^{(2)} \right \},
\end{equation}
where $|{\vec x}|^2= \sum_{i=1}^{7}\left(x^i\right)^2$. Clearly, (7.20)
is a smooth metric on the manifold
\begin{equation}  %7.21
R^7 \times M_2.
\end{equation}

The solution (7.20), (7.21) is an extension of the special
solution defined for $\tau \in (0, +\infty)$ to the
semi-interval $\tau \in (0,+ \infty]$. It may be interpreted in two
ways. First, we may say that the metric (7.20) describes an extension
of the 4-dimensional $R \times M_2$ cosmological solution to the case of
7-dimensional Euclidean time manifold.  On the other hand it may be also
considered as a spherically-symmetric ($O(7)$-symmetric) solution
with a curved time manifold $M_2$ (of negative curvature).

It seems likely that the procedure considered here  may be also
applied to the special solution with $C_1=\frac{\pi}{2}$,
$\varepsilon_1= -1$, $\varepsilon_2= +1$ for $N_1=6$, $N_2=3$ and for
other special solutions in $(1+8+2)$- and $(1+5+5)$-models.  We may also
obtain the "Milne-like" solution, when $(M_1,g^{(1)})$ is
the $6$-dimensional Lobachevsky space $(H^6,g(H^6))$.

%%%%%%%%%%%%%%%%%%%%%%%%%%%%%%%%%%%%%%%%%%%%%%%%%%%%%%%%%%%%%%%%%%%%%%%%%%%
%%%%%%%%%%%%%%%%%%%%%%%%%%%%%%%%%%%%%%%%%%%%%%%%%%%%%%%%%%%%%%%%%%%%%%%%%%%
%%%%%%%%%%%%%%%%%%%%%%%%%%%%%%%%%%%%%%%%%%%%%%%%%%%%%%%%%%%%%%%%%%%%%%%%%%%%
\section{Discussion}
\par

In the considered paper we have integrated the vacuum
Einstein equations for the model describing the
evolution of two Einstein spaces $M_1$ and $M_2$
with dimensions  $(N_1, N_2) = (6,3), (5,5), (8,2)$.
To our knowledge these are the first non-trivial
cosmological solutions describing the evolution of
more then one Einstein spaces of non-zero curvatures.

The Kasner-like behaviour near the singularity (for
$t_s \to +0$) is investigated.
The Kasner parameters (5.7), (5.8) are rational for
all considered three cases. We may consider the
following hypothesis: the Abel equation (2.20)
may be integrated by methods described in \cite{ZP,PZ}
(or by its extension) for the dimensions
satisfying
\begin{equation} %8.1
{R}(N_1, N_2) \equiv \sqrt{N_1 N_2 (N_1 + N_2 - 1)} = m \in {\bf Z}.
\end{equation}
($m$ is integer). In this (and only in this case) the Kasner parameters
(5.7), (5.8) are rational. (The relation (8.1) is satisfied
for $(N_1, N_2) = (13,13), (25,3), (25,25), (41,41)$ etc).

Here we also received some special solutions
to Einstein equations, i.e. "Milne-type" solutions for
arbitrary $n$ (see (4.2)), and two-parametric families
for $3 \leq n \leq 5$ (see (6.13)). We also obtained non-singular
solutions with topology $R^7 \times M_2$ (sec. 7).
(It may be shown that the non-singular solutions with
topology $R^{N_1 + 1} \times M_2^{N_2}$  exist also for
other dimensions $(N_1, N_2) = (5,5), (8,2)$.)

The considered in this paper solutions to the vacuum Einstein equations
are defined on the manifold $M$ of dimension  $D = $dim$M = 10, 11$.
These solutions satisfy the equations of motion
for $D = 10$ supergravity of superstring origin and
for $D = 11$, $N =1$ supergravity respectively. For certain manifolds
$M_1, M_2$ it is possible to use the obtained here solutions
for generation of other classical solutions
with non-zero matter fields (dilatonic, Kalb-Ramond etc)
in $D = 10, 11$ supergravities.
(This may  be done, for example, by using the duality
transformations.)

\begin{center}
{\bf Acknowledgments}
\end{center}
\par

The authors are grateful to K.A.Bronnikov, M.L.Fil'chenkov,
M.Yu.Konstantinov and D.-E.Liebscher for useful discussions.
\par
This work was supported  in part by the Russian Ministry of Science and
Russian Fund of Basic Sciences.

\pagebreak

Fig. 1. The graphical representation for the function
$2 {\Phi}(\tau, - \delta, C_1)$ (3.12) for $\delta = +1$ and
a) $C_1 > 0$, b) $C_1 = 0$, c) $C_1 < 0$, respectively.

\vskip 10mm

Fig. 2. The graphical representation for the function
$2 {\Phi}(\tau, - \delta, C_1)$ (3.12) for $\delta = -1$ and
a) $0 \leq C_1 < \frac{\pi}{2}$, b) $C_1 \geq \frac{\pi}{2}$.

\vskip 10mm

Fig. 3. The trajectories on the plane of scale factors $a_i = {a_i}(\tau),
i =1,2$, (corresponding to the solution (3.23)) with two negative
curvatures ($\varepsilon_1 = \varepsilon_2 = -1$). Here a) $\tau \in (1, +
\infty)$, for $C_1 \geq 0$; $\tau \in (1, {\tau_3}(C_1))$ for $C_1 < 0$;
b) $\tau \in ({\tau_2}(C_1), 1)$.

\vskip 10mm

Fig. 4. The trajectories on the plane of scale factors $a_i = {a_i}(\tau),
i =1,2$, for the case $\varepsilon_1 = + 1$,  $\varepsilon_2 = -1$,
(6-dimensional space has positive curvature and 3-dimensional space has
negative curvature). Here a) $\tau \in ({\tau_3}(C_1), + \infty)$ for
$C_1 < 0$; b) $\tau \in ({\tau_1}(C_1), {\tau_2}(C_1))$ for $C_1 \geq 0$.

\vskip 10mm

Fig. 5. The trajectories on the plane of scale factors $a_i = {a_i}(\tau),
i =1,2$, for the case $\varepsilon_2 = + 1$ (3-dimensional space has
positive curvature) and a) $\varepsilon_1 = +1$  (6-dimensional space has
positive curvature), $\tau \in (- \infty, {\tau_1}(C_1))$ for $C_1  >
   \frac{\pi}{2}$; b) $\varepsilon_1 = -1$ (6-dimensional space has
negative curvature), $\tau \in (- \infty, + \infty)$ for $C_1 \leq
   \frac{\pi}{2}$;  $\tau \in ({\tau_1}(C_1), + \infty)$ for $C_1 >
   \frac{\pi}{2}$.

\pagebreak

\end{document}